\documentclass[prd,showpacs,showkeys,superscriptaddress,preprintnumbers,nofootinbib]{revtex4}
\usepackage[latin9]{inputenc}
\usepackage{color}
\usepackage{amsmath}
\usepackage{amssymb}
\usepackage[unicode=true,
 bookmarks=false,
 breaklinks=false,pdfborder={0 0 1},backref=section,colorlinks=true]
 {hyperref}
\hypersetup{
 linkcolor=blue,urlcolor=blue,citecolor=red}

\makeatletter
\@ifundefined{textcolor}{}
{%
 \definecolor{BLACK}{gray}{0}
 \definecolor{WHITE}{gray}{1}
 \definecolor{RED}{rgb}{1,0,0}
 \definecolor{GREEN}{rgb}{0,1,0}
 \definecolor{BLUE}{rgb}{0,0,1}
 \definecolor{CYAN}{cmyk}{1,0,0,0}
 \definecolor{MAGENTA}{cmyk}{0,1,0,0}
 \definecolor{YELLOW}{cmyk}{0,0,1,0}
}

\usepackage{color}
\usepackage{graphicx}

\@ifundefined{textcolor}{}{
 \definecolor{BLACK}{gray}{0}
 \definecolor{WHITE}{gray}{1}
 \definecolor{RED}{rgb}{1,0,0}
 \definecolor{GREEN}{rgb}{0,1,0}
 \definecolor{BLUE}{rgb}{0,0,1}
 \definecolor{CYAN}{cmyk}{1,0,0,0}
 \definecolor{MAGENTA}{cmyk}{0,1,0,0}
 \definecolor{YELLOW}{cmyk}{0,0,1,0}
}

\makeatother

\begin{document}
\title{Double-logarithmic nonlinear electrodynamics }
\author{Ibrahim Gullu}
\email{ibrahim.gullu@emu.edu.tr}

\author{S. Habib Mazharimousavi}
\email{habib.mazhari@emu.edu.tr}

\affiliation{Department of Physics, Faculty of Arts and Sciences, Eastern Mediterranean
University, Famagusta, North Cyprus via Mersin 10, Turkey}
\date{\today }
\begin{abstract}
A new model of nonlinear electrodynamics named as \emph{``double-logarithmic''}
is introduced and investigated. The theory carries one dimensionful
parameter of the $\beta$ as Born-Infeld electrodynamics. It is shown
that the dual symmetry and dilatation (scale) symmetry are broken
in the proposed model. The electric field of a point-like charge is
derived for this model and it becomes non-singular at the origin and
by use of this electric field the static electric energy of a point
like charge is calculated. In the presence of an external magnetic
field the theory shows the phenomenon known as vacuum birefringence.
The refraction index of two polarizations, parallel and perpendicular
to the external magnetic induction field are calculated. The canonical
and symmetrical Belinfante energy-momentum tensors are obtained. Using
the causality and unitarity principles the regions where the theory
becomes causal and unitary are found. 
\end{abstract}
\maketitle

\section{Introduction}

It is known that there is a deep connection between non-linearity
and strong fields. When a classical field become enough strong it
invalidates the predictions get by the linear theory. There are examples
of classical fields in which the relation between non-linear effects
and strong fields has been applied. The very well known one is the
Born-Infeld (BI) electrodynamics \cite{Born,Born-Infeld}. In order
to solve the problem of singularity, namely infinite self-energy of
a point-like charge, which comes out in Maxwell's electrodynamics,
Born and Infeld introduced non-linearity to the classical electrodynamics.
BI electrodynamics solved this problem by introducing a dimensionful
parameter which gives an upper bound to the electric field of a point-like
charge. Moreover, BI type Lagrangians are used to understand the phenomena
of meson multiple production in the strong field regime \cite{Heisenberg},
and also the propagation of shock waves \cite{Taniuti}. Furthermore,
one-loop quantum corrections in quantum electrodynamics fixes the
Lagrangian of classical electrodynamics by non-linear terms \cite{Heisenberg-Euler,Schwinger,Adler}.
It is also possible to get non-linear electrodynamical effects in
strong gravity fields.

Besides the BI electrodynamics there are other type of non-linear
electrodynamics models. Some of them are known as logarithmic electrodynamics
\cite{Gaete-Neto}, exponential electrodynamics \cite{Hendi} and
$\arcsin$ electrodynamics \cite{Kruglov-1}. As well as, different
type of non-linear electrodynamics are considered in \cite{Kruglov-2,Kruglov-3,Kruglov-4,Kruglov-5,Kruglov-6}.

In this paper we introduce and analyze a new model of non-linear electrodynamics
which has significant properties like BI electrodynamics such as finite
electric field and electric field energy of a point-like charge. In
this model only one dimensional parameter, $\beta$, is introduced.

The layout of the paper is as follows. In Sec. II, we introduce the
model and show that the Maxwell's electrodynamics can be regained
with $\beta\rightarrow0$ limit. Also, the equation of motion is calculated
by use of which the Maxwell's equations are derived. Using the field
equations it is shown that the dual symmetry of the theory is broken.
The electric field of a point particle is calculated at the point
where the particle is located. In Sec. III, we calculate the speed
of the electromagnetic wave in presence of a constant and uniform
magnetic field and see the effect of vacuum birefringence. The Canonical
and Belinfante energy-momentum tensors, dilatation current and energy
of a point-like charge are obtained in Sec. IV. Sec. V is devoted
to the unitary and causality analysis of the theory. We end up with
a conclusion part which is Sec. VI.

In the continuation, we take $\hbar=c=\varepsilon_{0}=\mu_{0}=G=1$
and the Minkowski metric with mostly plus signature. The coordinates
are defined as $x^{\mu}=\left(t,r,\theta,\phi\right)$. Greek indices
run from $0$ to $3$ and Latin indices run from $1$ to $3$.

\section{The Model}

In the Maxwell's electrodynamics model, the Lagrangian is directly
proportional to $\mathcal{F}=\frac{1}{4}F_{\mu\nu}F^{\mu\nu}=\frac{\mathbf{B}^{2}-\mathbf{E}^{2}}{2},$
i.e., 
\begin{equation}
\mathcal{L}=-\mathcal{F}\text{.}\label{Maxwell}
\end{equation}
On the other hand, the most well known nonlinear theory i.e., BI theory,
proposes the following Lagrangian \cite{Born,Born-Infeld}
\begin{equation}
\mathcal{L}=b^{2}\left(1-\sqrt{1+\frac{\mathcal{F}}{b^{2}}-\frac{\mathcal{G}^{2}}{b^{4}}}\right)\label{BI}
\end{equation}
in which $\mathcal{G}=\frac{1}{4}F_{\mu\nu}\tilde{F}^{\mu\nu}=-\mathbf{B}\boldsymbol{\cdot}\mathbf{E.}$
The other well known nonlinear theory is the one proposed by Heisenberg
and Euler in 1936, \cite{Heisenberg-Euler,Heisenberg} with
\begin{equation}
\mathcal{L}=-\mathcal{F}-\beta\mathcal{F}^{2}-\gamma\mathcal{G}^{2}+...\text{.}\label{HE}
\end{equation}
Other than these historical models of the nonlinear electrodynamics
there are some recent models which have been intensively studied in
the literature. Among them exists the Maxwell's power law theory with
\cite{Hassaine,Gonzalez,Maeda}
\begin{equation}
\mathcal{L}=\alpha\left(-\mathcal{F}\right)^{s}\label{MPL}
\end{equation}
which was proposed by M. Hassaine and C. Martinez. More recently,
the $\arcsin$ model has been introduced by Kruglov with \cite{Kruglov-1}
\begin{equation}
\mathcal{L}=-\mathcal{F}-\frac{C}{\beta}\arcsin\left(\beta\mathcal{F}\right)+\frac{\gamma}{2}\mathcal{G}^{2}.\label{Kruglov}
\end{equation}
It is not difficult to see that for a single point charge the electric
field of the theories such as BI and Kruglov \cite{Kruglov-1} are
confined. However, although the electric field in BI theory is expressed
in terms of an elementary function, in the theories such as $\arcsin$
model such simple expression does not exist. Apart from the model
introduced in this manuscript there are few nonlinear electrodynamics
models which admits an exact solution for the electric field \cite{Mazharimousavi}.
Having the electric field confined and expressed in terms of an elementary
function are considered advantages of a theory on nonlinear electrodynamics.
This is because of its further applications.

Keeping in mind the finiteness and closed expression of the electric
field of a point charge, here, in this paper we introduce a new nonlinear
electrodynamics model given by 
\begin{equation}
\mathcal{L}=\frac{1}{2\beta}\left[\left(1-\mathcal{Y}\right)\ln\left(1-\mathcal{Y}\right)+\left(1+\mathcal{Y}\right)\ln\left(1+\mathcal{Y}\right)\right],\label{Lagrangian}
\end{equation}
in which, $\mathcal{Y}=\sqrt{-2\beta\mathcal{F}+\sigma\beta^{2}\mathcal{G}^{2}},$
with $\mathcal{F}$ and $\mathcal{G}$ the Maxwell invariants. We
shall refer to this as the \textit{double-logarithmic} nonlinear electrodynamics
model. Herein, the electromagnetic field and its dual are given by
$F_{\mu\nu}=\partial_{\mu}A_{\nu}-\partial_{\nu}A_{\mu}$ and $\tilde{F}^{\mu\nu}=\frac{1}{2}\epsilon^{\sigma\rho\mu\nu}F_{\sigma\rho}$
in which $A_{\mu}$ is the gauge potential and $\epsilon^{\alpha\beta\mu\nu}$
is completely antisymmetric tensor with the convention $\epsilon_{0123}=1$.
Furthermore, the coupling constant $\sigma$ can only take values
of $1,-1$ and $0$ and the Maxwell invariants should satisfy $-2\beta\mathcal{F}+\sigma\beta^{2}\mathcal{G}^{2}<1$
in order for $\mathcal{L}$ to be physically acceptable. Note that,
when $-2\beta\mathcal{F}+\sigma\beta^{2}\mathcal{G}^{2}<0$ the first
term and the second term in the Lagrangian (\ref{Lagrangian}) are
the complex conjugate of each other such that their addition becomes
real. On the other hand for $-2\beta\mathcal{F}+\sigma\beta^{2}\mathcal{G}^{2}=0$
(\ref{Lagrangian}) becomes zero too. The electromagnetic strength
tensor and its dual are explicitly given by 
\begin{equation}
F^{\mu\nu}=\left(\begin{array}{cccc}
0 & E^{1} & E^{2} & E^{3}\\
-E^{1} & 0 & B^{3} & -B^{2}\\
-E^{2} & -B^{3} & 0 & B^{1}\\
-E^{3} & B^{2} & -B^{1} & 0
\end{array}\right),\tilde{F}^{\mu\nu}=\left(\begin{array}{cccc}
0 & B^{1} & B^{2} & B^{3}\\
-B^{1} & 0 & -E^{3} & E^{2}\\
-B^{2} & E^{3} & 0 & -E^{1}\\
-B^{3} & -E^{2} & E^{1} & 0
\end{array}\right).\label{Electromagnetic_Strength}
\end{equation}

The $\beta\rightarrow0$ limit of (\ref{Lagrangian}) gives 
\begin{equation}
\lim_{\beta\rightarrow0}\mathcal{L}=-\mathcal{F}+\frac{1}{6}\left(2\mathcal{F}^{2}+3\mathcal{G}^{2}\sigma\right)\beta-\frac{1}{15}\left(4\mathcal{F}^{3}+5\mathcal{FG}^{2}\sigma\right)\beta^{2}+\mathcal{O}\left(\beta^{3}\right),\label{b_zero_limit}
\end{equation}
where the zeroth order term is the usual Maxwell's theory which can
be achieved when $\beta=0$. The first order correction to the Maxwell's
theory is $2\mathcal{F}^{2}+3\mathcal{G}^{2}\sigma$. For $\sigma=0$
the expansion becomes 
\begin{equation}
-\mathcal{F}+\frac{1}{3}\mathcal{F}^{2}\beta-\frac{4}{15}\mathcal{F}^{3}\beta^{2}+\mathcal{O}\left(\beta^{3}\right).\label{b_zero_limit_s_0}
\end{equation}
On the other hand, upon considering $-2\beta\mathcal{F}+\sigma\beta^{2}\mathcal{G}^{2}<1$,
$\lim_{\beta\rightarrow\infty}\mathcal{L}=0$.

\subsection{The Field Equations of the Model}

With the Lagrangian (\ref{Lagrangian}) we can write the following
action 
\begin{equation}
I=\int d^{4}x\sqrt{-\eta}\mathcal{L},\label{Action}
\end{equation}
in which $\eta$ is the determinant of $\eta_{\alpha\beta}$ the metric
of the Minkowski spacetime. Using the Euler-Lagrange equations 
\begin{equation}
\frac{\partial\mathcal{L}}{\partial A_{\beta}}-\partial_{\alpha}\frac{\partial\mathcal{L}}{\partial\partial_{\alpha}A_{\beta}}=0,\label{Euler_Lagrange}
\end{equation}
the field equations can be found as 
\begin{equation}
\partial_{\alpha}\left\{ \frac{\left(-F^{\alpha\beta}+\sigma\beta G\tilde{F}^{\alpha\beta}\right)}{\sqrt{-2\beta\mathcal{F}+\sigma\beta^{2}\mathcal{G}^{2}}}\ln\left(\frac{1+\sqrt{-2\beta\mathcal{F}+\sigma\beta^{2}\mathcal{G}^{2}}}{1-\sqrt{-2\beta\mathcal{F}+\sigma\beta^{2}\mathcal{G}^{2}}}\right)\right\} =0.\label{eom}
\end{equation}
The electric displacement field can be calculated, using the expression
$\mathbf{D}=\frac{\partial\mathcal{L}}{\partial\mathbf{E}},$ and
is given by 
\begin{equation}
\mathbf{D}=\frac{1}{2}\frac{\left(\overrightarrow{E}-\sigma\beta G\overrightarrow{B}\right)}{\sqrt{-2\beta\mathcal{F}+\sigma\beta^{2}\mathcal{G}^{2}}}\ln\left(\frac{1+\sqrt{-2\beta\mathcal{F}+\sigma\beta^{2}\mathcal{G}^{2}}}{1-\sqrt{-2\beta\mathcal{F}+\sigma\beta^{2}\mathcal{G}^{2}}}\right).\label{electric_displacement}
\end{equation}
We can write (\ref{electric_displacement}) in terms of the electric
permittivity tensor $\varepsilon_{ij}$ which is 
\begin{equation}
\varepsilon_{i}^{j}=\varepsilon\left(\delta_{i}^{j}+\sigma\beta B_{i}B^{j}\right),\label{electric_permittivity}
\end{equation}
where 
\begin{equation}
\varepsilon\equiv\frac{1}{2\sqrt{-2\beta\mathcal{F}+\sigma\beta^{2}\mathcal{G}^{2}}}\ln\left(\frac{1+\sqrt{-2\beta\mathcal{F}+\sigma\beta^{2}\mathcal{G}^{2}}}{1-\sqrt{-2\beta\mathcal{F}+\sigma\beta^{2}\mathcal{G}^{2}}}\right).\label{Epsilon}
\end{equation}
Then the electric displacement field (\ref{electric_displacement})
takes the following form 
\begin{equation}
D^{j}=\varepsilon\left(\delta_{i}^{j}+\sigma\beta B_{i}B^{j}\right)E^{i}=\varepsilon_{i}^{j}E^{i}.\label{electric_displacement_1}
\end{equation}
Once $\sigma$ is taken $0$ the electric displacement field becomes
$D^{j}=\tilde{\varepsilon}E^{j}$ where $\tilde{\varepsilon}=\frac{1}{2\sqrt{-2\beta\mathcal{F}}}\ln\left(\frac{1+\sqrt{-2\beta\mathcal{F}}}{1-\sqrt{-2\beta\mathcal{F}}}\right)$.
The magnetic field is given by $\mathbf{H}=-\frac{\partial\mathcal{L}}{\partial\mathbf{B}}$,
and after taking the derivative of (\ref{Lagrangian}) with respect
to magnetic field we obtain 
\begin{equation}
\mathbf{H}=\frac{\mathbf{B}+\sigma\beta\mathcal{G}\mathbf{E}}{2\sqrt{-2\beta\mathcal{F}+\sigma\beta^{2}\mathcal{G}^{2}}}\ln\left(\frac{1+\sqrt{-2\beta\mathcal{F}+\sigma\beta^{2}\mathcal{G}^{2}}}{1-\sqrt{-2\beta\mathcal{F}+\sigma\beta^{2}\mathcal{G}^{2}}}\right).\label{magnetic_field}
\end{equation}
Introducing the inverse magnetic permeability tensor 
\begin{equation}
\left(\mu^{-1}\right)_{i}^{j}=\varepsilon\left(\delta_{i}^{j}-\sigma\beta E_{i}E^{j}\right)\label{magnetic_permeability}
\end{equation}
one writes the magnetic field 
\begin{equation}
H^{j}=\varepsilon\left(B^{j}+\sigma\beta\mathcal{G}E^{j}\right)=\left(\mu^{-1}\right)_{i}^{j}B^{i},\label{magnetic_field_1}
\end{equation}
or equivalently 
\begin{equation}
B^{i}=\mu_{j}^{i}H^{j},\label{magnetic_induction_field}
\end{equation}
where $B^{i}$ is the magnetic induction field. Without the Maxwell
invariant $\mathcal{G}$, the magnetic field becomes $B^{i}=\tilde{\varepsilon}H^{i}.$
In the first two pair of Maxwell's equations (\ref{electric_displacement_1})
and (\ref{magnetic_field_1}) the appearance of the electric permittivity
tensor $\varepsilon_{i}^{j}$ and magnetic permeability tensor $\mu_{i}^{j},$
respectively, signs a medium with complicated properties.

The equations of motion (\ref{eom}) can be rewritten in the form
of the first pair of the Maxwell's equations upon using (\ref{electric_displacement})
and (\ref{magnetic_field}). For $\beta=0$ in (\ref{eom}) we get

\begin{equation}
\mathbf{\triangledown}\cdot\mathbf{D}=0\label{Maxwell_1}
\end{equation}
and for $\beta=i$ we get the second Maxwell equation 
\begin{equation}
-\frac{\partial\mathbf{D}}{\partial t}+\mathbf{\triangledown}\times\mathbf{H}=0.\label{Maxwell_2}
\end{equation}
The second pair of the Maxwell's equations can be obtained by the
Bianchi identity $\partial_{\mu}\tilde{F}^{\mu\nu}=0$. First we set
$\nu=0$ in the Bianchi identity which give
\begin{equation}
\mathbf{\triangledown}\cdot\mathbf{B}=0,\label{Maxwell_3}
\end{equation}
and then $\nu=j$ to get 
\begin{equation}
\frac{\partial}{\partial t}\mathbf{B}+\left(\mathbf{\triangledown}\times\mathbf{E}\right)=0.\label{Maxwell_4}
\end{equation}
In order to see the dual symmetry of the theory we take the dot product
of the magnetic field (\ref{magnetic_field}) and the electric displacement
field (\ref{electric_displacement}) i.e., 
\begin{equation}
\mathbf{D\cdot H}=\varepsilon^{2}\mathbf{E\cdot B}\neq\mathbf{E\cdot B}\label{dual_symmetry}
\end{equation}
which states that the dual symmetry is broken for $\sigma=0$ and
$\sigma\neq0,$ unlike the BI theory. When we take $\beta\rightarrow0$
limit we arrive at classical electrodynamics and the dual symmetry
is recovered

\begin{equation}
\mathbf{D\cdot H}=\mathbf{E\cdot B}.\label{dual_symmetry_of_Maxwell}
\end{equation}

\subsection{Electrostatics}

Let us consider electrostatics for which $\mathbf{B}=\mathbf{H}=\mathbf{0}$
and find the electric field at the origin of the point-like charged
particle namely at $r=0$. The equation of point-like charge is 
\begin{equation}
\mathbf{\triangledown}\cdot\mathbf{D}_{0}=q\delta\left(\overrightarrow{r}\right)\label{eq_of_point_like_charge}
\end{equation}
in which $q$ is the electric charge. The latter admits a solution
given by 
\begin{equation}
\mathbf{D}_{0}=D_{0}\hat{r}=\frac{q}{4\pi r^{2}}\hat{r},\label{solution_of_point-like_charge}
\end{equation}
where 
\begin{equation}
D_{0}=\frac{1}{2\sqrt{-2\beta\mathcal{F}}}\ln\left(\frac{1+\sqrt{-2\beta\mathcal{F}}}{1-\sqrt{-2\beta\mathcal{F}}}\right)E_{0}.\label{solution_D_0}
\end{equation}
When there is no magnetic field the Maxwell invariant $\mathcal{F}$
becomes $\mathcal{F}=-\frac{1}{2}E_{0}^{2}$ such that 
\begin{equation}
D_{0}=\frac{1}{2\sqrt{\beta}}\ln\left(\frac{1+\sqrt{\beta}E_{0}}{1-\sqrt{\beta}E_{0}}\right).\label{solution_D_0_1}
\end{equation}
Furthermore, the equation takes the following form 
\begin{equation}
\frac{1}{2\sqrt{\beta}}\ln\left(\frac{1+\sqrt{\beta}E_{0}}{1-\sqrt{\beta}E_{0}}\right)=\frac{q}{4\pi r^{2}},\label{solution_of_point_like_charge_1}
\end{equation}
and finally the solution for the electric field is given by 
\begin{equation}
E_{0}=\frac{\tanh\left[\frac{q\sqrt{\beta}}{4\pi r^{2}}\right]}{\sqrt{\beta}}.\label{Electric_of_point_like}
\end{equation}
Here we can define unitless variable to check $r\rightarrow0$ limit
by defining 
\begin{equation}
x\equiv\frac{2\pi r^{2}}{q\sqrt{\beta}},\quad y\equiv E_{0}\sqrt{\beta},\label{unitless_variables}
\end{equation}
then $r\rightarrow0$ limit becomes $x\rightarrow0$ limit. With these
unitless variables the electric field equation (\ref{Electric_of_point_like})
is rewritten as 
\begin{equation}
y=\tanh\left(\frac{1}{2x}\right),\label{Electric_field_unitless_variables}
\end{equation}
whose $x\rightarrow0$ limit gives $y\rightarrow1$. Therefore the
maximum value for the electric field at the origin, where the charged
particle is placed, is given by 
\begin{equation}
E_{\text{max}}=\sqrt{\frac{1}{\beta}},\label{E_max}
\end{equation}
which shows the finitness of the electric field at the location of
the point particle as BI electrodynamics.

\section{Vacuum Birefringence}

In this part we are going to discuss the effect of vacuum birefringence
which relates the phase velocity to the polarization of the electromagnetic
wave. Hence, let us take an external magnetic induction field $\mathbf{B}_{0}=\left(B_{0},0,0\right)$
which is uniform and constant together with the plane electromagnetic
wave $\left(\mathbf{e},\mathbf{b}\right)$ as 
\begin{equation}
\mathbf{e}=\mathbf{e}_{0}\exp\left[-i\left(\omega t-kz\right)\right],\quad\mathbf{b}=\mathbf{b}_{0}\exp\left[-i\left(\omega t-kz\right)\right]\label{emw}
\end{equation}
propagating in the $z$-direction. As a result the total electromagnetic
field become $\mathbf{E}=\mathbf{e}$ and $\mathbf{B}=\mathbf{b}+\mathbf{B}_{0}$.
Since we are interested in strong magnetic induction field the amplitudes
of the electromagnetic wave $e_{0},\;b_{0}$ become small compared
to the magnetic induction field, that is $e_{0,\;}b_{0}\ll B_{0}$.
After linearizing the equations (\ref{electric_displacement}, \ref{magnetic_field})
the electric permittivity and magnetic permeability tensors can be
found as 
\begin{equation}
\varepsilon_{ij}=\varepsilon\left(\eta_{ij}+\sigma\beta\eta_{i1}\eta_{j1}B_{0}^{2}\right),\label{electric_permittivity_1}
\end{equation}
and 
\begin{equation}
\mu_{ij}=\mu\eta_{ij}\quad\mu=\varepsilon^{-1}.\label{magnetic_permrability_1}
\end{equation}
From Maxwell's equations one can obtain the following wave equation
\begin{equation}
\partial_{j}^{2}E_{i}-\mu\varepsilon_{ij}\partial_{t}^{2}E_{j}-\partial_{i}\partial_{j}E_{j}=0.\label{wave_eq}
\end{equation}
Choosing the polarization of the electric field parallel to the external
magnetic field i.e., $\mathbf{e}_{0}=e_{0}\left(1,0,0\right)$ and
solving the wave equation (\ref{wave_eq}) we find 
\begin{equation}
\mu\varepsilon_{11}\omega^{2}=k^{2}.
\end{equation}
The index of refraction $n_{\parallel}=\sqrt{\mu\varepsilon_{11}}$
can be found as 
\begin{equation}
n_{\parallel}=\sqrt{\varepsilon^{-1}\varepsilon\left(\eta_{11}+\sigma\beta\eta_{11}\eta_{11}B_{0}^{2}\right)}=\sqrt{1+\sigma\beta B_{0}^{2}}.
\end{equation}
If the polarization is chosen such that the electromagnetic wave is
perpendicular to the external induction field i.e., $\mathbf{e}_{0}=e_{0}\left(0,1,0\right)$
then (\ref{wave_eq}) gives 
\begin{equation}
\mu\varepsilon_{22}\omega^{2}=k^{2},
\end{equation}
and the index of refraction become 
\begin{equation}
n_{\perp}=\sqrt{\mu\varepsilon_{22}}=\sqrt{\varepsilon^{-1}\varepsilon\left(\eta_{22}+\sigma\beta\eta_{21}\eta_{21}B_{0}^{2}\right)}=1.
\end{equation}
After all, the phase velocity depends on the polarization and takes
the value $v_{||}=\frac{1}{n_{||}}$ for parallel polarization i.e.,
$\mathbf{e}\Vert\mathbf{B}_{0}$ and $v_{\perp}=1$ ($c=1$) for perpendicular
polarization i.e., $\mathbf{e}\perp\mathbf{B}_{0}$. This shows the
effect of vacuum birefringence. On the other side, if we get rid of
the Maxwell invariant $\mathcal{G}$ by setting $\sigma=0$ the effect
of vacuum birefringence cancel out.

\section{The Energy-Momentum Tensor}

In this section we derive the energy-momentum tensor to find the dilatation
current and the energy of point-like charge. The general expression
of the canonical energy-momentum tensor is given as 
\begin{equation}
T_{\nu}^{\mu\left(C\right)}=\left(\partial_{\nu}A_{\alpha}\right)\frac{\partial\mathcal{L}}{\partial\left(\partial_{\mu}A_{\alpha}\right)}-\delta_{\nu}^{\mu}\mathcal{L},
\end{equation}
which upon using (\ref{Lagrangian}) one obtains 
\begin{equation}
T_{\nu}^{\mu\left(C\right)}=\varepsilon\left(\partial_{\nu}A_{\alpha}\right)\left(-F^{\mu\alpha}+\sigma\beta\mathcal{G}\tilde{F}^{\mu\alpha}\right)-\delta_{\nu}^{\mu}\mathcal{L}.
\end{equation}
While this tensor is conserved $\partial_{\mu}T^{\left(C\right)\mu\nu}=0$
it is not symmetric and gauge-invariant. Therefore, we have to obtain
the symmetric Belinfante tensor \cite{Coleman-Jackiw} by introducing
\begin{equation}
T_{\nu}^{\mu\left(B\right)}=T_{\nu}^{\mu\left(C\right)}+\partial_{\beta}\chi_{\nu}^{\beta\mu}.
\end{equation}
Herein, 
\begin{equation}
\chi_{\beta\mu\nu}=\frac{1}{2}\left[\Pi_{\beta\sigma}\left(\Sigma_{\mu\nu}\right)^{\sigma\rho}-\Pi_{\mu\sigma}\left(\Sigma_{\beta\nu}\right)^{\sigma\rho}-\Pi_{\nu\sigma}\left(\Sigma_{\beta\mu}\right)^{\sigma\rho}\right]A_{\rho},\label{chi}
\end{equation}
where 
\begin{equation}
\Pi^{\beta\sigma}=\frac{\partial\mathcal{L}}{\partial\left(\partial_{\beta}A_{\sigma}\right)}=\varepsilon\left(-F^{\beta\sigma}+\sigma\beta G\tilde{F}^{\beta\sigma}\right),\label{Pi}
\end{equation}
and the generators of Lorentz transformations $\Sigma_{\mu\alpha}$
have the matrix elements 
\begin{equation}
\left(\Sigma_{\mu\alpha}\right)_{\sigma\rho}=\eta_{\mu\sigma}\eta_{\alpha\rho}-\eta_{\mu\rho}\eta_{\alpha\sigma}.\label{Sigma}
\end{equation}
Using (\ref{chi}) and (\ref{Sigma}) one achieves 
\begin{equation}
\chi_{\nu}^{\phantom{\nu}\beta\mu}=\Pi^{\beta\mu}A_{\nu},\label{chi_1}
\end{equation}
which shows $\chi_{\beta\mu\nu}=-\chi_{\mu\beta\nu}$ that gives rise
to $\partial^{\mu}\partial^{\beta}\chi_{\beta\mu\nu}=0$. Therefore,
the symmetrical Belinfante tensor is conserved. From (\ref{chi_1})
and (\ref{Pi}) we find 
\begin{equation}
\partial_{\beta}\chi_{\nu}^{\phantom{\nu}\beta\mu}=\varepsilon\left(-F^{\beta\mu}\partial_{\beta}A_{\nu}+\sigma\beta G\tilde{F}^{\beta\mu}\partial_{\beta}A_{\nu}\right),
\end{equation}
where $\partial_{\beta}\Pi^{\beta\mu}=0$ upon (\ref{eom}). Finally,
the Belinfante tensor can be written as follows 
\begin{equation}
T_{\nu}^{\mu\left(B\right)}=-\varepsilon\left(-F^{\mu\alpha}F_{\nu\alpha}+\sigma\beta GF_{\nu\alpha}\tilde{F}^{\mu\alpha}\right)+\delta_{\nu}^{\mu}\mathcal{L}.\label{B1}
\end{equation}
The trace of the Belinfante tensor (\ref{B1}) is found to be 
\begin{equation}
T^{B}=\frac{2\mathcal{F}}{\sqrt{-2\beta\mathcal{F}+\sigma\beta^{2}\mathcal{G}^{2}}}\ln\left(\frac{1+\sqrt{-2\beta\mathcal{F}+\sigma\beta^{2}\mathcal{G}^{2}}}{1-\sqrt{-2\beta\mathcal{F}+\sigma\beta^{2}\mathcal{G}^{2}}}\right)-\frac{2}{\beta}\ln\left(1+2\beta\mathcal{F}-\sigma\beta^{2}\mathcal{G}^{2}\right),\label{trace_of_em}
\end{equation}
which will be needed to see the dilatation current. For the linear
electrodynamics $\beta$ is taken to zero limit and in this limit
the trace of the energy-momentum tensor becomes zero. In this model
(\ref{Lagrangian}) non-zero trace of energy-momentum tensor is appeared.

\subsection{Dilatation Current}

In this part the dilatation current is calculated to see the scale
symmetry of the model. The modified dilatation current is defined
as \cite{Coleman-Jackiw} 
\begin{equation}
D_{\mu}^{B}=x^{\alpha}T_{\mu\alpha}^{B}+V_{\mu},\label{modified_dilatation}
\end{equation}
where the field-virial is given by 
\begin{equation}
V_{\mu}=\Pi_{\alpha\beta}\left[\delta_{\mu}^{\alpha}\delta_{\rho}^{\beta}-\left(\Sigma_{\mu}^{\alpha}\right)_{\rho}^{\beta}\right]A_{\rho}.\label{viral}
\end{equation}
It can be shown easily that $V_{\mu}=0$, and the modified dilatation
current becomes $D_{\mu}^{B}=x^{\alpha}T_{\mu\alpha}^{B}$. The divergence
of the dilatation current is 
\begin{equation}
\partial^{\mu}D_{\mu}^{B}=T^{B}.\label{divergence_of_current}
\end{equation}
As a result the scale symmetry is broken due to the dimensional parameter
$\beta$. Although the Maxwell's theory is conformal symmetrical,
in BI theory both the scale and conformal symmetries are broken.

\subsection{Energy Density of Pure Electric Field}

The energy density, $\rho_{E}=T_{00}^{B}$ for pure electric energy
($\mathbf{B}=0$) is

\begin{equation}
T_{00}^{B}=-\frac{1}{2\beta}\ln\left(1-\beta E^{2}\right)\label{energy_density}
\end{equation}
which is positive as it is expected. The total electric energy of
a point charge is given by $\mathcal{E}=\int\rho_{E}dV$ where $\rho_{E}$
is the energy density derived above. Hence, one writes 
\begin{equation}
\mathcal{E}=-\frac{2\pi}{\beta}\int_{0}^{\infty}\ln\left(1-\beta E^{2}\right)r^{2}dr.\label{Total_electric_energy_1}
\end{equation}
In terms of the dimensionless variables, this equation becomes 
\begin{equation}
\mathcal{E}=\frac{2\pi}{\beta}\left(\frac{q\sqrt{\beta}}{4\pi}\right)^{\frac{3}{2}}\int_{0}^{\infty}\frac{\ln\left(\cosh x\right)}{x^{5/2}}dx,\label{Total_electric_energy_2}
\end{equation}
where we have used $x=\frac{r_{0}^{2}}{r^{2}}$ and $r_{0}=\sqrt{\frac{q\sqrt{\beta}}{4\pi}.}$
Our numerical calculation yields 
\begin{equation}
\mathcal{E}=0.101\frac{q^{2}}{r_{0}}.\label{Total_electric_energy}
\end{equation}
Following \cite{Nature}, if we assume the rest mass of the electron
is electromagnetic one obtains 
\begin{equation}
mc^{2}=0.101\frac{q^{2}}{r_{0}}\label{b_value}
\end{equation}
which in turn provides estimations for $\beta.$

\section{The Causality and Unitarity}

In this part we will analyze the unitarity and causality of the particle
spectrum of the theory. If the theory satisfies the following inequalities
\cite{Shabad-Usov} 
\begin{eqnarray}
\mathcal{L}_{\mathcal{F}}\leq0 & ,\quad\mathcal{L}_{\mathcal{FF}}\geq0, & \quad\mathcal{L}_{\mathcal{GG}}\geq0,\nonumber \\
\mathcal{L}_{\mathcal{F}}+2\mathcal{FL}_{\mathcal{FF}}\leq0, &  & \quad2\mathcal{FL}_{\mathcal{GG}}-\mathcal{L}_{\mathcal{F}}\geq0.\label{unitary_and_causality}
\end{eqnarray}
it will be non-tachyonic, in which the particles has a group velocity
lower than the speed of light and ghost free which means that the
particles have a positive kinetic energy or a lower bound for that
energy. Here the subscripts refers to the partial derivatives of the
Lagrangian density with respect to the indicated invariant.

In our model derivatives of the Lagrangian density (\ref{Lagrangian})
are 
\begin{equation}
\mathcal{L}_{\mathcal{F}}=-\frac{\tanh^{-1}\left(\sqrt{\beta\Delta}\right)}{\sqrt{\beta\Delta}},\label{L_F}
\end{equation}
\begin{equation}
\mathcal{L}_{\mathcal{FF}}=-\frac{\beta\tanh^{-1}\left(\sqrt{\beta\Delta}\right)}{\left(\beta\Delta\right)^{\frac{3}{2}}}+\frac{1}{\Delta-\beta\Delta^{2}},\label{L_FF}
\end{equation}

\begin{equation}
\mathcal{L}_{\mathcal{G}}=\frac{\sigma\beta\mathcal{G}\tanh^{-1}\left(\sqrt{\beta\Delta}\right)}{\sqrt{\beta\Delta}},\label{L_G}
\end{equation}
and 
\begin{equation}
\mathcal{L}_{\mathcal{GG}}=\frac{-\mathcal{G}^{2}\beta^{3}\sigma^{2}\sqrt{\beta\Delta}}{\left(\beta\Delta\right)^{3/2}\left(-1+\beta\Delta\right)}+\frac{2\beta^{2}\sigma\mathcal{F}\left(1-\beta\Delta\right)\tanh^{-1}\left(\sqrt{\beta\Delta}\right)}{\left(\beta\Delta\right)^{3/2}\left(-1+\beta\Delta\right)}.\label{L_GG}
\end{equation}
The other two inequalities of the principles (\ref{unitary_and_causality})
involve 
\begin{equation}
\mathcal{L}_{\mathcal{F}}+2\mathcal{FL}_{\mathcal{FF}}=-\frac{\sigma\beta\mathcal{G}^{2}\tanh^{-1}\left(\sqrt{\beta\Delta}\right)}{\sqrt{\beta\Delta}\Delta}+\frac{2\mathcal{F}}{\Delta-\beta\Delta^{2}},\label{L_F_plus_L_FF}
\end{equation}
and 
\begin{equation}
2\mathcal{F}\mathcal{L}_{\mathcal{GG}}-\mathcal{L}_{\mathcal{F}}=-\frac{2\mathcal{FG}^{2}\beta^{3}\sigma^{2}\sqrt{\beta\Delta}}{\left(\beta\Delta\right)^{3/2}\left(-1+\beta\Delta\right)}+\frac{4\beta^{2}\sigma\mathcal{F}^{2}\left(1-\beta\Delta\right)\tanh^{-1}\left(\sqrt{\beta\Delta}\right)}{\left(\beta\Delta\right)^{3/2}\left(-1+\beta\Delta\right)}+\frac{\tanh^{-1}\left(\sqrt{\beta\Delta}\right)}{\sqrt{\beta\Delta}},\label{F_L_GG_minusL_F}
\end{equation}
where we have defined a new variable $\Delta,$ given by 
\begin{equation}
\Delta\equiv-2\mathcal{F}+\sigma\beta\mathcal{G}^{2}\label{Delta}
\end{equation}
to have simpler relations and assumed $\beta>0$. 

\subsection{Causality and Unitarity of the Model}

The first condition of (\ref{unitary_and_causality}) states that
\begin{equation}
\frac{\tanh^{-1}\left(\sqrt{\beta\Delta}\right)}{\sqrt{\beta\Delta}}\geq0,\label{first_cond}
\end{equation}
while the second one implies
\begin{equation}
\frac{1}{1-\left(\beta\Delta\right)}\geq\frac{\tanh\left(\sqrt{\beta\Delta}\right)}{\sqrt{\beta\Delta}}.\label{second_cond}
\end{equation}
Using (\ref{L_GG}), the third condition of (\ref{unitary_and_causality})
can be expressed as 
\begin{equation}
2\sigma\mathcal{F}\left(1-\beta\Delta\right)\tanh^{-1}\left(\sqrt{\beta\Delta}\right)\leq\mathcal{G}^{2}\beta\sigma^{2}\sqrt{\beta\Delta}.\label{third_cond}
\end{equation}
From the fourth condition of (\ref{unitary_and_causality}) we get
\begin{equation}
\frac{2\mathcal{F}}{1-\beta\Delta}\leq\frac{\sigma\mathcal{\beta G}^{2}\tanh^{-1}\left(\sqrt{\beta\Delta}\right)}{\sqrt{\beta\Delta}},\label{fourth_cond}
\end{equation}
and finally the last one of (\ref{unitary_and_causality}) represents
\begin{equation}
\left(1-\frac{4\beta^{2}\sigma\mathcal{F}^{2}}{\beta\Delta}\right)\tanh^{-1}\left(\sqrt{\beta\Delta}\right)\geq\frac{2\mathcal{FG}^{2}\beta^{3}\sigma^{2}}{\sqrt{\beta\Delta}\left(-1+\beta\Delta\right)}.\label{fifth_cond}
\end{equation}
Beside these conditions there is one more condition which comes from
the reality of $\ln\left(1-y\right)$ and the reality of the trace
of energy-momentum tensor (\ref{trace_of_em}), that is 
\begin{equation}
-2\beta\mathcal{F}+\sigma\beta^{2}\mathcal{G}^{2}<1,\label{reality_cond}
\end{equation}
which can be retyped as
\begin{equation}
\beta\Delta<1.\label{reality_cond_1}
\end{equation}
We can discuss the unitarity of the model however the discussion depends
on the choice of $\sigma$. Put $\sigma=0$ and $\sigma=-1$ cases
by and concentrate on $\sigma=1$ case.

\subsubsection{$\sigma=1$ case:}

We have chosen $\beta$ to be positive as well as $\sigma=1$ and
with the reality condition (\ref{reality_cond_1}) $\Delta$ has to
be positive. The first condition (\ref{first_cond}) is automatically
satisfied for this choices. The second condition (\ref{second_cond})
can be rewritten as 
\begin{equation}
1-\beta\Delta-\frac{\left(\beta\Delta\right)^{\frac{1}{2}}}{\tanh^{-1}\left(\sqrt{\beta\Delta}\right)}\leq0\label{second_cond_1}
\end{equation}
which is always satisfied as long as $\Delta>0$. The rest of the
conditions (\ref{third_cond}), (\ref{fourth_cond}) and (\ref{fifth_cond})
are analyzed by plotting two dimensional graphics. As a result the
(\ref{third_cond}) and (\ref{fourth_cond}) are satisfied whereas
the last condition (\ref{fifth_cond}) is satisfied for a confined
magnetic field which depends on the value of $\beta$. 

In the next sections we discuss the unitary of electric part and magnetic
part of the model (\ref{Lagrangian}) separately by taking $\sigma=0$.

\subsubsection{The Unitarity of the Electric Part:}

In this part we are going to analyses the unitarity of the theory
considering only the electric part of it by taking $\sigma=0$ and
$B=0$. The first (\ref{first_cond}) and the fifth (\ref{fifth_cond})
unitarity conditions give 
\begin{equation}
\frac{\tanh^{-1}\left(\sqrt{\beta}E\right)}{\sqrt{\beta}E}\geq0\label{unitary_electric}
\end{equation}
which states that $E<\frac{1}{\sqrt{\beta}}$. Since the maximum value
for the electric field of a point charge is $E_{\text{max}}=\frac{1}{\sqrt{\beta}}$
these conditions are satisfied. The third condition (\ref{third_cond})
is also satisfied since $\mathcal{L}_{\mathcal{GG}}=0$. The fourth
condition (\ref{fourth_cond}) states $E<\frac{1}{\sqrt{\beta}}$
which is the same condition as the first (\ref{first_cond}) and the
fifth (\ref{fifth_cond}) ones. The second condition (\ref{second_cond})
can be written as 
\begin{equation}
\frac{\sqrt{\beta}E}{1-\beta E^{2}}\geq\tanh^{-1}\left(\sqrt{\beta}E\right)\label{unitary_electric_1}
\end{equation}
which is always satisfied for $\sqrt{\beta}E<1$.

\subsubsection{The Unitarity of the Magnetic Part}

In order to analyze the unitarity of the magnetic part we choose $\sigma=0$,
$E=0$. The first (\ref{first_cond}) and second (\ref{second_cond})
conditions are satisfied due to the nice property of $\tanh^{-1}\left(ix\right)=i\tanh^{-1}\left(x\right)$.
The third condition (\ref{third_cond}) is automatically gratified.
The fourth condition (\ref{fourth_cond}) can be written in terms
of the magnetic field as 
\begin{align}
-\frac{1}{1+\beta B^{2}} & \le0,
\end{align}
which is always satisfied. Since the fifth (\ref{fifth_cond}) condition
reduces to the first one it is also fulfilled. As a result, the pure
magnetic case of (\ref{Lagrangian}) is unitary.

\section{Conclusion:}

We have introduced a new nonlinear electrodynamics model, the \emph{double-logarithmic}
model. The model contains both of the Maxwell invariants $\mathcal{F}$
and $\mathcal{G}$ and it carries one dimensionful parameter $\beta$.
After finding the field equations of the theory we calculated the
electric field of a point like charge and we showed that at the origin--the
location of the charge--it takes a finite value, that is $E_{\text{max}}=\frac{1}{\sqrt{\beta}}$,
and is not singular. We showed also that, in presence of a magnetic
field the model admits the effect of vacuum birefringence in which
the phase velocities of electromagnetic wave depend on the polarizations.
This effect disappears once the Maxwell invariant $\mathcal{G}$ is
disposed. We obtain the canonical and symmetric Belinfante energy-momentum
tensors in order to calculate the dilatation current. We showed that
the dilatation symmetry is broken due to the dimensional parameter
$\beta$. Moreover, the self-energy of a point-like charge is calculated.
We also discuss the unitarity and causality of the model and conclude
that the general model has a unitary region for $\beta>0$ and $\sigma=1$
as long as the magnetic field is confined. The unitarity and causality
of electric and magnetic parts of the model are discussed separately
and it is shown that not only the electric part fulfills all the conditions
to be unitary and causal with the following constraints $E<\frac{1}{\sqrt{\beta}}$
and $\beta>0$ but also the magnetic part gratifies the unitary and
causality conditions.

\end{document}